\documentclass[aps,pra,a4paper,showpacs,twocolumn]{revtex4}
\usepackage{amssymb}

\usepackage{amsmath}
\usepackage{bm}
\usepackage{graphicx}
\usepackage{ulem}

\begin{document}
\title{Field-free molecular alignment induced by elliptically polarized laser pulses: non invasive 3 dimensional characterization}

\author{E. Hertz,$^{1}$ D. Daems,$^{2}$ S. Gu\'erin,$^{1}$ H.R. Jauslin,$^{1}$ B. Lavorel,$^{1}$ and O. Faucher$^{1}$}
\affiliation{$^{1}$Institut Carnot de Bourgogne, UMR 5209 CNRS-Universit\'e de Bourgogne,
BP 47870, 21078 Dijon, France\\
$^{2}$Quantum Information and Communication, Ecole Polytechnique, Universit\'e Libre
de Bruxelles, 1050 Brussels, Belgium}

\begin{abstract}

An investigation of field-free molecular alignment produced by elliptically polarized laser pulses is reported. Experiments are conducted in CO$_2$ at room temperature. A non invasive all-optical technique, based on the cross defocusing of a probe pulse, is used to measure the alignment along two orthogonal directions that is sufficient to provide a 3 dimensional characterization. The field-free molecular alignment produced by a laser of elliptical polarization is in good agreement in terms of amplitude and shape with theoretical predictions. It turns out to be almost equivalent to the superposition of the effects that one would obtain with two individual cross-polarized pulses. The investigation highlights notably the occurrence of field-free two-direction alignment alternation for a suitably chosen degree of ellipticity. The analogy between this specific ellipticity and the well known "magic angle" used in time resolved spectroscopy to prevent rotational contributions is discussed.  

\end{abstract}
\maketitle

\section{Introduction}

The possibility of controlling the angular distribution of molecules in space has focused considerable interest in the very recent years \cite{Stap, Seideman}. It is now well established that molecules exposed to intense non resonant linearly polarized laser pulses tend to align along the direction of the electric field. In the long pulse limit (i.e. when the pulse duration is long compared to the rotational period), the molecule is trapped adiabatically in pendular states and the original isotropic state is recovered upon turn off. In contrast, a short laser pulse leaves the system in a coherent superposition of rotational states. The rotational wavepacket thus produced revives periodically in time leading to the occurrence of post-pulse transient alignment observed experimentally by several groups \cite{Vrakking,VRenard,Dooley}. A large variety of processes depends on the relative molecular orientation. Spatial manipulation of molecules can be exploited in order to control these processes \cite{Larsen, Itatani, Zeidler},  to gain some insight into their orientation dependence \cite{Litvinyuk, Pinkham}, or just to prepare a system in which the process is simpler to analyse than with a sample of randomly oriented molecules. Non adiabatic alignment is better suited for further applications since it provides a sample of field-free aligned molecules. In this regard, techniques designed to control the dynamics of post-pulse alignment are particularly valuable. Various strategies, mainly based on pulse shaping techniques, have been proposed. By appropriately tailoring the laser pulse, optimization of alignment \cite{Bisgaard}, switching of alignment \cite{Spanner,RenardPRA2005}, and control of the rephasing period \cite{RenardPRA2004} have been demonstrated. 
Pulse shaping is not the only approach. The set of control parameters can be extended for instance to laser polarization which offers promising potentialities in terms of spatial manipulation of molecules. Elliptically polarized laser pulses have been applied in the adiabatic regime to establish 3D-alignment~\cite{Larsen3D} and -orientation~\cite{Tanji3D} of planar molecules. In the non-adiabatic regime, we have reported in a recent Letter \cite{Daems} the alternation of post-pulse alignment of CO$_{2}$ along two orthogonal directions using a specific ellipticity. In the present paper, we extend this analysis by investigating the alignment of CO$_{2}$ produced with different laser polarizations. We show in particular that the ellipticity of Ref.~\cite{Daems} is related to  the magic angle used in the context of time resolved spectroscopy~\cite{Magic}. In the following, the alignment is measured along two orthogonal directions with a non invasive optical technique which relies on the cross defocusing of a probe beam due to the spatial distribution of aligned molecules. A first "pump" pulse focused into a molecular gas sample induces post-pulse alignment. The degree of alignment is larger at the center of the laser focus than at the edge as a result of the intensity distribution. This spatial distribution of aligned molecules results in a spatial gradient of the refractive index that is read by a  weak time-delayed "probe" pulse. By measuring the increase of the probe beam size in the far field region as a function of the delay, the field-free molecular alignment can be determined \cite{VRenard2005}.  In the present work, the degree of alignment along two perpendicular directions is measured using two different probe polarizations. Since the alignment along the third direction (or along any other direction) can be deduced from the two precedent ones, the measurement provides a complete 3 dimensional characterization of the molecular alignment in space. We investigate both amplitude and dynamics of transient alignment versus laser ellipticity.

\section{Experimental setup}

\begin{figure}[tbph]
\centerline{\includegraphics[scale=0.4]{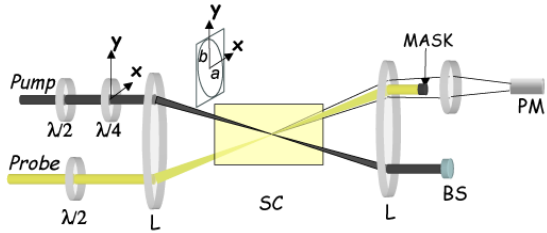}}
\caption{(Color online) Experimental set-up. L : lens, $\lambda$/2 : half waveplate, $\lambda$/4 : quarter waveplate, SC : static cell, BS : Beam Stop, PM: Photomultiplier. }
\label{Fig1}
\end{figure}

The experimental set-up of the cross defocusing technique is succinctly depicted in Fig.~\ref{Fig1}. The laser apparatus is based on a chirped pulse amplified Ti:sapphire femtosecond laser. The system delivers 100 fs pulses of  wavelength around 800 nm at a 1 KHz repetition rate. The probe pulse of weak intensity is produced through a reflection of the beam on a glass plate. The time delay between pump and probe pulse is scanned by a corner cube retro-reflector mounted on a motorized linear stage. The degree of ellipticity of the pump field is controlled by a half waveplate combined with a quarter waveplate. Depolarization induced by reflective optics has been minimized by positioning the set of waveplates at the very end of the set-up. 

The electric field is written as
\begin{equation}
\label{vectE}
\vec{E}(t)=\Lambda(t) \left[a\cos(\omega t)\vec{x}+b\sin(\omega t)\vec{y}\right],
\end{equation}
\noindent 
with 0$\leq$$a$$\leq$1 and 0$\leq$$b$$\leq$1 respectively the relative field amplitudes along  the neutral axis $\vec{x}$ and $\vec{y}$  of the quarter waveplate ($a^{2}$+$b^{2}$=1),  and $\Lambda(t)$  the pulse envelope. The degree of ellipticity is adjusted by tuning the half waveplate axes. It ranges in the present investigation from linear polarization along the $\vec{y}$ axis ($a^{2}$=0)  to circular ($a^{2}=1/2$). $\vec{x}$ and $\vec{y}$ correspond thus respectively to the minor and major axis of the ellipse. The probe polarization is set alternatively along these two axes by means of a second half waveplate. The pump and probe beams are focused by the same lens of 30 cm focal length. They cross at a small angle ($\sim2^{\circ}$) in a long static cell filled with CO$_2$ at room temperature and pressure of 4 $\times$ 10$^{4}$ Pa. The measured beam waists are 30~$\mu$m  with a \textit{M}$^{2}$ factor \cite{Siegman} below 1.3. At the exit of the cell, the pump beam is blocked whereas the modification of the spatial profile of the probe beam is detected by using a coronograph as follows. The central part of the probe beam is blocked by a mask with a diameter adjusted to produce a signal close to zero for isotropic distribution of molecules in the cell. When molecular alignment occurs, a refractive index gradient is created and the size of the probe beam in the far field region increases. The light passing around the mask is detected by a photomultiplier and the signal sampled by a boxcar is stored in a computer. We have shown recently \cite{VRenard2005} that the present cross defocusing technique provides a signal proportional to $\left[\langle \cos^2 \theta\rangle(t)-1/3\right]^2$,  where $\langle \cos^2 \theta\rangle$ is the quantum and thermally averaged value that characterizes  the extent of alignment and $\theta$ the angle between the molecular axis and the probe polarization \cite{Stap, Seideman}. By selecting the latter either in the $x$ or $y$ direction, the quantities $\left[\langle \cos^2 \theta_{x} \rangle(t)-1/3\right]^2$ or $\left[\langle \cos^2 \theta_{y}\rangle(t)-1/3\right]^2$ can be measured. The alignment  $\langle \cos^2 \theta_{z} \rangle(t)$  along the $z$  axis (orthogonal to the ellipse's plane) can be then deduced through the geometric relation

\begin{equation}
\label{sumcos2}
\sum_{i=x,y,z}\langle \cos^2 \theta_{i} \rangle(t)=1.
\end{equation}
\noindent 
A complete three dimensional characterization of the post-pulse alignment is thus in principle achieved. It should be noted that the present cross defocusing technique measures independently the variation of the refractive index along the $x$ or the $y$ axis as opposed to the birefringence technique \cite{VRenard} that measures the difference. The cross defocusing has thus the advantage to allow alignment measurement along different axes. Furthermore, it should be noted that the signal produced by the cross defocusing technique is proportional to the degree of alignment produced at the center of the laser beam exposed to the peak intensity. It does not arise therefore from a standard spatial averaging over the intensity distribution of the profile of the laser beam. The model of the cross defocusing  technique indeed intrinsically takes into account the spatial profile of the laser beam through the refractive index gradient \cite{VRenard2005}. Experiments have been conducted at a pump peak intensity of 25 TW/cm$^{2}$ insuring that ionization remains low. The cross defocusing technique is in fact sensitive to ionization. This property has been recently used in order to provide quantitative measurements of ionization probabilities in N$_2$ \cite{Loriot}.

\section{Results and discussion}

\begin{figure}[tbph]
\centerline{\includegraphics[scale=1.5]{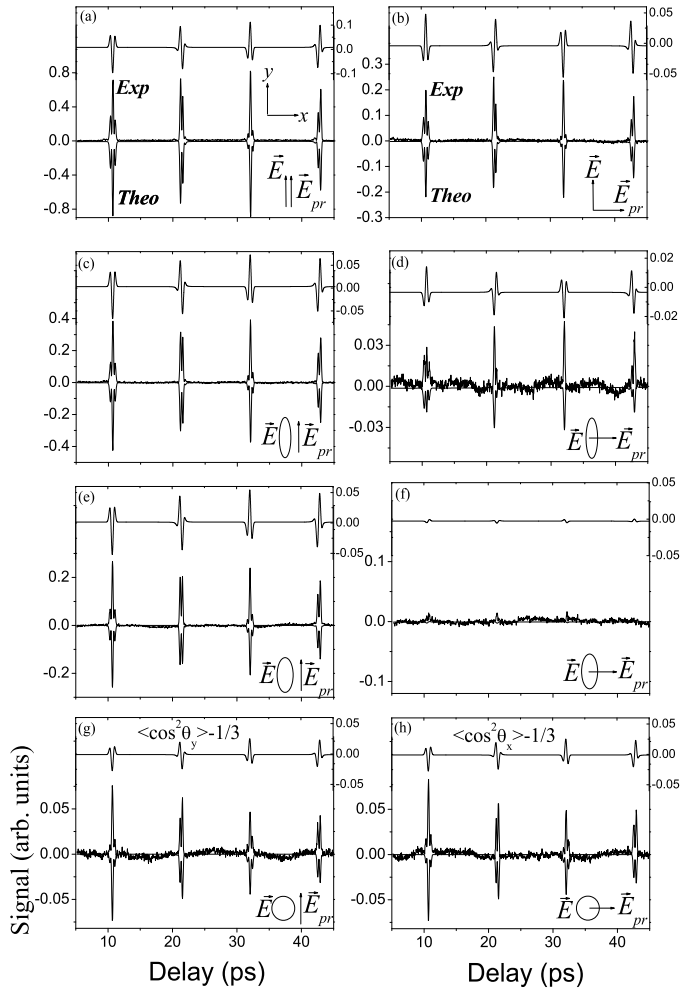}}
\caption{Cross defocusing signal (\textit{Exp}) measured with the probe field polarized along the $y$-axis (left column) and along the $x$-axis (right column) with the calculation (\textit{Theo}) shown reversed for different ellipticities defined by $a^{2}\simeq$0 (a, b), 1/4 (c,d), 1/3 (e, f), and 1/2 (g, h). The corresponding observables $\langle \cos^2 \theta_{y}\rangle(t)-1/3$ and $\langle \cos^2 \theta_{x}\rangle(t)-1/3$ are depicted in the upper part of each panel. The alignment along the third axis (orthogonal to the ellipse's plane), i.e. $\langle \cos^2 \theta_{z}\rangle(t)-1/3$, can be deduced from Eq.~(\ref{sumcos2}).}
\label{Fig2}
\end{figure}

\begin{figure}[tbph]
\centerline{\includegraphics[scale=1]{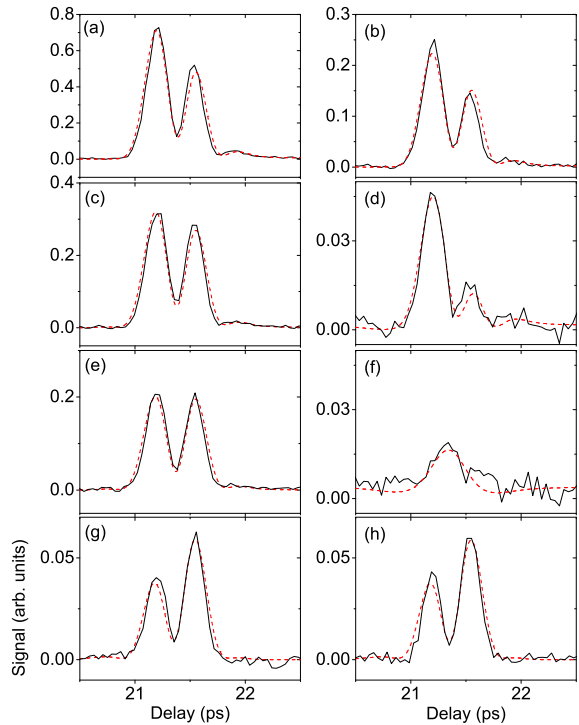}}
\caption{(Color online) Cross defocusing signal (full line) measured with the probe field polarized along the $y$-axis (left column) and along the $x$-axis (right column) for $a^{2}\simeq$0 (a, b), 1/4 (c,d), 1/3 (e, f), and 1/2 (g, h). The calculations, shown as dashed lines, are multiplied by an adjusted scaling factor.}
\label{Fig3}
\end{figure}

\begin{figure}[tbph]
\centerline{\includegraphics[scale=1]{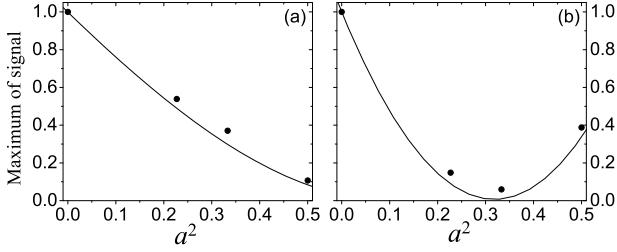}}
\caption{Maximum of signal at the first revival as a function of $a^{2}$ with the probe field polarized along the $y$-axis (a) and along the $x$-axis (b). The measurements are normalized with respect to the linear polarization $a^{2}$=0. Calculation is shown as a solid line.}
\label{Fig4}
\end{figure}

Fig. \ref{Fig2}  shows experimental measurements obtained for a set of pump ellipticities characterized by $a^{2}\simeq$[0, 1/4, 1/3, 1/2], ranging from linear (along $y$) to circular. 
The left column displays experimental curves (\textit{Exp}) with the probe field polarized along the $y$ axis, i.e producing a signal proportional to $\left[\langle \cos^2 \theta_{y}\rangle(t)-1/3\right]^2$, while the right column shows signals with the probe polarized along the $x$ axis, i.e. proportional to $\left[\langle \cos^2 \theta_{x}\rangle(t)-1/3\right]^2$. It should be noted that all data have been recorded with the detection sensitivity kept constant so that the amplitude of traces can be directly compared. For each ellipticity, the observables $\left[\langle \cos^2 \theta_{i}\rangle(t)-1/3\right]$ $i$=$x,y$ are inserted in the upper part of each panel. These observables have been obtained through the  numerical simulation of the time-dependent Schr\"{o}dinger equation described in Ref.~\cite{Daems}. The signals predicted by the theory (\textit{Theo}), corresponding to the convolution of $\left[\langle \cos^2 \theta_{i}\rangle(t)-1/3\right]^2$ $i$=$x,y$ with the probe pulse intensity envelope, are shown as mirror images of the experimental ones. For a direct comparison with the experiment, theoretical curves are multiplied by the same scaling factor that gives the possibility to follow the modification of the amplitude with the ellipticity. The experimental and calculated signals can be thus directly compared in terms of shape and amplitude. We emphasize nevertheless that the degree of alignment is determined through the shape of the pump-probe signal and not through the amplitude (although a calibration procedure is possible~\cite{VRenardPRA}). The temporal shape of the quantity $\left[\langle \cos^2 \theta\rangle(t)-1/3\right]^2$ is indeed very sensitive to the degree of alignment [i.e. to small changes in $\langle \cos^2 \theta\rangle(t)$] due to the occurrence of permanent alignment that modifies the asymmetry of the revivals~\cite{VRenard}. As shown in Fig.~\ref{Fig2}, the modification of the transient shapes with the ellipticity is statisfactorily reproduced even for signals of strongly reduced amplitude. In particular, the second revival (around 21~ps) exhibits a notable modification of asymmetry as expected. We focus in  Fig.~\ref{Fig3} on this revival to compare its shape with the predicted one. The scaling factor is now adjusted for each ellipticity since our concern is limited here to the shape of the revivals. Except for Fig.~\ref{Fig3}(f) of too small amplitude, all the other revivals display the predicted shape corroborating the expectation of Fig.~\ref{Fig2}. 
This outcome supports the validity of $\langle \cos^2 \theta_{i}\rangle(t)$ $i$=$x,y$ used in the calculation and shown in upper part of Fig.~\ref{Fig2}. These quantities can be thus used in  Eq.~(\ref{sumcos2}) to deduce the alignment along the $z$ axis. The variation of the signal amplitude is expected to be reproduced as well. In a general view (Fig.~\ref{Fig2}), as the ellipticity increases, the signal (\textit{Exp}) along the $y$ axis decreases continuously while it exhibits a minimum along the $x$ axis for $a^{2}$=1/3 (Fig.~\ref{Fig2}~(f)) as expected by the theory (\textit{Theo}). This feature, predicted theoretically (see Ref.~\cite{Daems}, Fig.~3),  is therefore confirmed experimentally in the present work. We have depicted in Fig. \ref{Fig4} the signal at the peak of the first revival (around 11 ps) for the different ellipticities of this work compared to the theory. 
Apart from small deviations, the global change is satisfactorily reproduced. For instance from Fig.~\ref{Fig2}~(a) to Fig.\ref{Fig2}~(b) the amplitude of the experimental signal is reduced by a factor about 4 (more precisely 3.6$\pm0.6$ \cite{ratio}). These curves have been obtained with the pump field linearly polarized along the $y$ axis. The $x$ and $z$ directions are thus equivalent, i.e. $\langle \cos^2 \theta_{x} \rangle(t)=\langle \cos^2 \theta_{z} \rangle(t)$. According to Eq.(\ref{sumcos2}), the relation between the signals measured along the $x$ and $y$ direction is then given by $\langle \cos^2 \theta_{x} \rangle(t)-1/3=-\left[\langle \cos^2 \theta_{y} \rangle(t)-1/3\right]/2$. The factor $-1/2$  (squared) is thus satisfactorily supported by the experimental data. The shape of the signals is also expected to be the same as observed experimentally. It has been reported recently (see Appendix B of Ref.\cite{Rouzee}) that alignment produced with an elliptically polarized laser pulse can be evaluated in the intermediate field regime \cite{IFR} by considering the outcome as the superposition of effects corresponding to two cross-polarized laser pulses in phase quadrature. We will show that this property allows one to interpret the ellipticity dependence of the alignment and provides particular straightforward predictions for the simulations depicted in Fig. \ref{Fig2}. For a linearly polarized pump field, we can label $A_{\parallel}$ and $A_{\perp}$ the post-pulse alignment respectively measured along and perpendicularly to the polarization direction. They are given by the following expressions
\begin{subequations}
\begin{eqnarray}
\label{Aparra}
A_{\parallel}=\left[\langle \cos^2 \theta_{y}\rangle(t)-1/3\right]_{a^{2}=0},\\
\label{Aperp}
A_{\perp}=\left[\langle \cos^2 \theta_{x}\rangle(t)-1/3\right]_{a^{2}=0}=-\frac{1}{2}\times A_{\parallel}
\end{eqnarray}
\end{subequations}
and are displayed respectively in the upper parts of Fig.~\ref{Fig2}~(a) and Fig.~\ref{Fig2}~(b).

For any degree of ellipticity, the laser field in the ($x$-$y$) plane can be written as the combination of linearly polarized fields along the $x$ and $y$ axis with a $\pi/2$-phase shift and weighting factors $a$ and $b$ [Eq.~(\ref{vectE})]. The alignment measured along the $y$ and $x$ axes considering the superposition of the effects of these fields is thus

\begin{subequations}
\begin{eqnarray}
\label{parraperpa}
\left(\langle \cos^2 \theta_{y}\rangle(t)-1/3\right)_{a^{2}}&\approx&\left(b^2  A_{\parallel}+a^2  A_{\perp}\right)\notag\\
&\approx&\left(1-3 a^2/2\right) A_{\parallel}\\
\label{parraperpb}
\left(\langle \cos^2 \theta_{x}\rangle(t)-1/3\right)_{a^{2}}&\approx&\left(a^2  A_{\parallel}+b^2  A_{\perp}\right)\notag\\
&\approx& \frac{3 a^2-1}{2} A_{\parallel}.
\end{eqnarray}
\end{subequations}
Through these two relations, the quantities $\langle \cos^2 \theta_{i=x,y}\rangle(t)-1/3$ shown in the upper panels of Fig.~\ref{Fig2} can be inferred. $a^2$=1/4  (Figs.~\ref{Fig2}(c) and \ref{Fig2}(d)) gives for instance $\langle \cos^2 \theta_{y}\rangle(t)-1/3\approx 5 A_{\parallel}/8$ and  $\langle \cos^2 \theta_{x}\rangle(t)-1/3 \approx - A_{\parallel}/8$. The $x$-axis alignment exhibits as expected a significant decrease without attaining nevertheless its minimum. The latter is expected to arise according to Eq.~(\ref{parraperpb}) for $a^2$=1/3. Experimental results shown in Fig.~\ref{Fig2}(f) yield a value $\langle \cos^2 \theta_{x}\rangle(t)\thickapprox 1/3$ (i.e. almost no alignment along the $x$-axis) corroborating this prediction. The specific ellipticity leading to this result exhibits an analogy with the so-called "magic angle". The latter is used notably in time resolved spectroscopy techniques in order to isolate the isotropic response~\cite{Magic}. Rotational coherence spectroscopy of excited states produces for instance a signal $I_{\parallel}$ when the pump and probe polarization are parallel and $I_{\perp}$  for crossed polarization. The relation between these two signals is $I_{\parallel}$=$-2 I_{\perp}$. When the probe pulse is set at the magic angle $\theta_{m}$=$\arctan\sqrt 2\approx54.735^{\circ}$ with respect to the pump, the detected signal does not exhibit any rotational coherence because these two effects cancel out ($I$=$I_{\parallel}\cos^{2}\theta_{m}+ I_{\perp} sin^{2}\theta_{m}$ =$ I_{\parallel}/3+2I_{\perp}/3$=0). In the present work, the $x$ and $y$ axis field components contribute also to $\langle \cos^2 \theta_{x}\rangle(t)-1/3$ in antiphase with a weighting factor two (see Eq.~(\ref{Aperp})). The ellipticity characterized by $a^2$=1/3 ($b^2$=2/3), obtained when the electric field before the quarter waveplate is set at the magic angle with respect to the $x$ neutral axis, confers twice the intensity along the $y$ axis, producing hence no alignment along the  $x$-axis, as observed. According to Eq.~(\ref{sumcos2}), $\langle \cos^2 \theta_{x}\rangle(t)\thickapprox-1/3$ implies  $\langle \cos^2 \theta_{x}\rangle(t)-1/3=- \left[\langle \cos^2 \theta_{z}\rangle(t)-1/3\right]$ signifying for this specific ellipticity a balanced alignment along the {$z$- and $y$-axis \cite{Daems}. The quasi-isotropic angular distribution along the $x$-axis  constitutes thus a clear signature of optimum alternation of alignment. This remarkable feature can be also inferred from the interaction Hamiltonian

\begin{equation}
\label{hamilt}
V_\text{int}=-\frac{\bigtriangleup\alpha}{4}\Lambda^{2}(t)\left(\left(1-2a^{2}\right) \cos^{2}\theta_{y}+a^{2}\sin^{2}\theta_{z}\right).
\end{equation}
The choice $a^{2}$=1/3 meets the criteria $1-2a^{2}$= $a^{2}$, imparting thus a symmetric role to $\theta_{y}$ and $\theta_{z}$. The resulting two directions alignment alternation could be used for instance to generate a 3D molecular switch. 
The results shown Fig.~\ref{Fig2}(g) and Fig.~\ref{Fig2}(h) are also particularly interesting. Circular polarization is known to produce planar delocalization in the polarization plane ($x$, $y$)  and alignment orthogonally to this plane, i.e. along the $z$ axis. The alignment along the $x$- and $y$-axes is thus expected to be the same which is verified by experimental observations. Using this feature combined with Eq.~(\ref{sumcos2}), the alignment along the $z$ axis can be deduced via the relation  $\langle \cos^2 \theta_{x}\rangle(t)-1/3=-1/2 \left(\langle \cos^2 \theta_{z}\rangle(t)-1/3\right)$. The alignment arising along the $z$-axis can be thus easily evaluated since it corresponds to the one measured along the $x$- or $y$-axis with a scaling factor $-1/2$. In view of applications, the alignment produced with circular polarization features notable advantages. A number of processes, like high harmonic generation or double ionization are indeed strongly minimized with circular polarization. If one wants to investigate these processes with samples of aligned molecules, it seems more suited to induce alignment with laser fields of circular polarization rather than linear. The measurements are then not hampered by any detrimental background signal associated to the "alignment pulse".


\section{Summary and conclusion}

We have investigated the field-free molecular alignment of linear molecules induced by elliptically polarized laser pulses. A 3D-angular characterization of the alignment is carried out by means of a pump-probe technique based on the cross defocusing of a weak probe pulse. The technique provides a signal proportional to $\left[\langle \cos^2 \theta\rangle(t)-1/3\right]^2$ with $\theta$ the angle between the molecular axis and the probe polarization. Through a theoretical analysis, this signal allows  a determination of the quantity $\langle \cos^2 \theta\rangle(t)$ that characterizes the extent of alignment along the probe polarization axis. In the present work, the quantities $\langle \cos^2 \theta_{x} \rangle(t)$  and $\langle \cos^2 \theta_{y} \rangle(t)$ along two perpendicular directions $x$ and $y$ are determined from separate measurements using different probe polarizations. Since the alignment $\langle \cos^2 \theta_{z} \rangle(t)$ along the third direction $z$ can be deduced from the two precedent ones via the geometric relation $\sum_{i=x,y,z}\langle \cos^2 \theta_{i} \rangle(t)=1$, the method provides a 3 dimensional characterization of the alignment. We emphasize that the degree of alignment along the $x$ or $y$ axis (i.e. $\langle \cos^2 \theta_{x} \rangle(t)$ or $\langle \cos^2 \theta_{y} \rangle(t)$) is determined by comparing the shape of the experimental transient revivals with the calculated ones. A small change of the degree of alignment (through for instance a variation of the laser intensity or the ellipticity) leads indeed to significant modifications of the shape of the signal. Thus, when the shape of the experimental signal is in good agreement with the calculated one, we can conclude that the numerical value of $\langle \cos^2 \theta\rangle(t)$ characterizes accurately the degree of alignment of the molecules in the experiment. The values $\left[\langle \cos^2 \theta_{i}\rangle(t)-1/3\right]^2$ ($i=x$, $y$) can be also in principle determined in absolute through a calibration with atoms of known non linear refractive index \cite{VRenardPRA}. However, in the present case it seems difficult to use this method in order to provide a 3 dimensional characterization of the alignment. The quantities $\langle \cos^2 \theta_{x} \rangle(t)$ and $\langle \cos^2 \theta_{y} \rangle(t)$ need to be known to determine $\langle \cos^2 \theta_{z} \rangle(t)$ and it is rather difficult to extract them directly from the signal because: (i) the signal corresponds to the convolution of the probe pulse with $\left[\langle \cos^2 \theta_{i}\rangle(t)-1/3\right]^2$ ($i=x, y$) so that the analysis would require a prior deconvolution of the signal, (ii) the experimental signal does not provide the sign of $\langle \cos^2 \theta_{i} \rangle(t)$ and finally (iii) this method is probably less precise since based on the comparison between two separate atomic and molecular signals. For all these reasons, it is more reliable to compare the shape of signals with the theory to determine the degree of alignment. In the present investigation, the pronounced modifications of the alignment in terms of shape and amplitude as a function of the laser ellipticity are in good agreement with theoretical predictions. The analysis points out a specific ellipticity leading to the alternation of alignment along two perpendicular axes as reported previously. An interpretation based on the alignment produced with elliptical polarization as the superposition of alignment produced by two linearly polarized laser fields is verified. Based on this simplified analysis, the specific ellipticity leading to the alternation of alignment is analogous to the implementation of the so called "magic angle". Possible benefits of the use of non-linearly polarized laser pulses in the frame of molecular alignment are discussed. The present work opens up the way to field-free molecular alignment produced with fields of temporally controlled polarization. In this context, the versatility of the detection technique used in this paper will be particularly valuable.

\acknowledgments 
The authors acknowledge financial support by the Conseil R\'egional de Bourgogne, the CNRS, the ACI "Photonique",
and a Marie Curie European Reintegration Grant within the 6th European Community RTD Framework Programme. Electronic address: edouard.hertz@u-bourgogne.fr.


\begin{references}
\bibitem{ratio} Each transient of the two experimental traces has been fitted with a scaling factor as a free parameter. The ratio has been evaluated by comparing the results.
\bibitem{IFR} The intermediate field regime corresponds to the regime well below intrinsic saturation regime but beyond the perturbative regime (for which no significant field-induced transfer of rotational population occurs).
\end{references}
\end{document}